\PassOptionsToPackage{usenames,dvipsnames,svgnames}{xcolor}
\documentclass[9pt,twocolumn,twoside]{pnas-new_dp}

\templatetype{pnasbriefreport} % Choose template 
\setboolean{displaywatermark}{false}
\usepackage{ctable}
\usepackage[font=sc,labelfont=sc, labelsep=none]{caption}
\usepackage{booktabs}
\usepackage{comment}
\usepackage{here}

\usepackage{siunitx}
\usepackage{sansmath}

\title{Collaborative knowledge exchange promotes innovation}
\author[a,b,1%\orcid{0000-0002-9053-5599}
]{Tomoya Mori}
\author[a%\orcid{0000-0003-3290-9721}
]{Jonathan Newton} 
\author[c%\orcid{0000-0001-7243-4383}
]{Shosei Sakaguchi}

\affil[a]{Institute of Economic Research, Kyoto University. Yoshida-Honmachi, Sakyo-Ku, Kyoto, Kyoto 606-8501, Japan.}
\affil[b]{Research Institute of Economy, Trade and Industry, 11th floor, Annex, Ministry of Economy, Trade and Industry 1-3-1, Kasumigaseki Chiyoda-ku, Tokyo 100-8901, Japan.}
\affil[c]{Faculty of Economics, The University of Tokyo, 7-3-1, Hongo, Bunkyo-ku, Tokyo
113-0033, Japan.}

\leadauthor{Mori} 

\authorcontributions{Author contributions:Conceptualization: T.M., J.N.; Funding acquisition: T.M; Theoretical analysis: J.N., T.M.; Empirical analysis: S.S., T.M.; Writing; J.N., T.M.
}
\authordeclaration{The authors declare no conflict of interest.}
\correspondingauthor{\textsuperscript{1}To whom correspondence should be addressed. E-mail: mori@\@kier.kyoto-u.ac.jp}


\dates{This manuscript was compiled on \today}

\begin{document}

%%%% WORD COUNTING %%%%
%TC:ignore
% \section*{Word Counts}

% This section is \textit{not} included in the word count.

% \detailtexcount{main}
%TC:endignore
%%%% WORD COUNTING (END) %%%%

\begin{abstract}
%Considering collaborative innovation in patent development, 
Considering collaborative patent development,
we provide micro-level evidence
%of knowledge spillovers 
for innovation through exchanges of differentiated knowledge.
Knowledge embodied in a patent is proxied by word pairs appearing in its abstract, while novelty is measured by the frequency with which these word pairs have appeared in past patents. 
Inventors are assumed to possess the knowledge associated with patents in which they have previously participated.
We find that collaboration by inventors with more mutually differentiated knowledge sets is likely to result in patents with higher novelty.
\end{abstract}

\maketitle
\setcounter{page}{1}

\thispagestyle{firststyle}
\ifthenelse{\boolean{shortarticle}}{\ifthenelse{\boolean{singlecolumn}}{\abscontentformatted}{\abscontent}}{}

\dropcap{H}umans are a collaborative species \cite{tomasello2007shared, tomasello2010ape, tomasello2003makes} and bring this collaborative nature with them to the workplace and to the laboratory. Collaboration within teams has been increasingly important within academic research \cite{wuchty2007increasing} and patent development \cite{Singh-Fleming-MS2010,Singh-et-al-SMJ2016}, and has long been an essential part of other creative endeavors such as the production of Broadway musicals \cite{uzzi2005collaboration}. 
A necessary condition for successful collaboration is successful coordination on roles and tasks \cite{AngusNewton2015PLOS, newton2017shared}, which can be complicated when the characteristics of one's collaborative partners are imperfectly understood \cite{rusch2019collaboration}.
Hence, collaboration requires shared knowledge among collaborators. 
However, for collaboration to be creative, it also requires differentiated knowledge \cite{Berliant-Fujita-IER2008}.
%Hence, for collaboration to both occur and be creative, it requires both shared knowledge and differentiated knowledge. 
There is, for example, evidence that the successful production of Broadway musicals involves teams that comprise both people who have previously collaborated and people who have not previously collaborated \cite{uzzi2005collaboration, bel2015team}. 

%Using Japanese patent data, we consider collaboration in the process of knowledge creation itself, 

Using Japanese patent data and adopting a measure of novelty based on word combinations that appeared in the patent \cite{Watzinger-et-al-CEPR2021}, we find a strong positive relationship between collaborators' mutual knowledge differentiation and the novelty of their output.
%, other things being equal.
%By proxying the novelty of output by that of word combinations appeared in the patent \cite{Watzinger-et-al-CEPR2021}, we find that collaborations of inventors with more mutually differentiated knowledge likely result in patents with higher novelty, other things being equal.

\section*{Results}
$\bf W_p$ represents the set of distinct word pairs in patent $p$'s abstract. (Hereafter, a bold capital letter expresses a set, and the corresponding italic letter its cardinality.)
The \emph{novelty} of a word pair at a given point in time is measured by the likelihood of its appearance in patents filed in the past \cite{Watzinger-et-al-CEPR2021}. Specifically, the novelty $n_{wt}$ of word pair $w$ at time $t$ is the ratio of (i) the sum of $W_p = |\bf W_p|$ over all patents $p$ filed at dates up to and including $t$, to (ii) the number of these patents that include word pair $w$.
We measure patent novelty by the average novelty of word pairs in its abstract, $\frac{1}{W_p}\sum_{w\in {\bf W}_p} n_{wt_p}$, where $t_p$ is the patent's filing time. 

We consider collaborative aspects of patent development by focusing on the productivity per inventor pair, following \cite{Berliant-Fujita-IER2008}.
${\bf H}_p$ is the set of all inventors who participated in patent $p$, while ${\bf M}_p\equiv \{(i,j): i,j,\in {\bf H}_p, i\neq j\}$ is the set of pairs of such inventors.
The \emph{average pairwise-contribution to the patent's novelty} is given by
\begin{equation*}
    n_{p}=\frac{1}{M_p W_p}\sum_{w\in {\bf W}_p} n_{wt_p}\,.
\end{equation*}%$n_p=\sum_{w\in \bf W_p} n_{wt_p}$, 

Denoting by ${\bf G}_{it}$ the set of patents inventor $i$ participated in at time $t$, define $i$'s \emph{knowledge} at $t$ by ${\bf K}_{it} = \cup_{\tau < t}\cup_{p\in {\bf G}_{i\tau}} {\bf W}_p$ and its novelty by $k_{it} = \sum_{w\in \mathbf{K}_{it}} n_{wt}$. 

Inventor pair $\{i,j\}$ has total knowledge $\mathbf K_{ijt} = \mathbf{K}_{it} \cup \mathbf K_{jt}$, with novelty $k_{ijt} = \sum_{w\in \mathbf K_{ijt}} n_{wt}$, and inventor $i$'s
 \emph{differentiated knowledge} relative to $j$ is
$\mathbf{K}^D_{ijt} = \mathbf K_{it}\backslash \mathbf K_{jt}$, with novelty $k^D_{ijt} = \sum_{w\in \mathbf K^D_{ijt}} n_{wt}$.

Knowledge differentiation between $\{i,j\}$ is evaluated by the geometric mean of their respective differentiated-knowledge shares in the union of their knowledge,
\begin{equation*}
    s_{ijt} = \sqrt{k^D_{ijt} k^D_{jit}}/k_{ijt}\in [0,0.5]\,.
\end{equation*}
%Let ${\bf H}_p$ and $\bold M_p\equiv \{(i,j)\::\: i,j,\in {\bf H}_p, i\neq j\}$ represent the sets of inventors and all pairs of inventors, respectively, participated in patent $p$.
Their average in patent $p$,
\begin{equation*}
    s_p = \frac{1}{M_p}\sum_{(i,j)\in\mathbf{M}_p}  s_{ijt_p}
\end{equation*}
measures knowledge differentiation in $p$. 
We focus on patents with $s_p>0$, since $s_p= 0$ implies no knowledge exchange as inventors can be indexed so that ${\bf K}_1\subseteq {\bf K}_2 \cdots \subseteq {\bf K}_{H_p}$. 

We estimate the effect of $s_p$ on $n_p$ by the model:
\begin{align}
n_{p}= \beta_0 &+ \beta_1 s_p + \cdots + \beta_m s_p^m \nonumber\\
    &+ \gamma \bar{K}_p+ \delta M_p + f_p + \varphi_p + \tau_p + \varepsilon_p\,, \label{eq:regression}
\end{align}
controlling for average knowledge size 
$\bar{K}_p \equiv \frac{1}{H_p}\sum_{i\in {\bf H}_p} K_{it_p}$, inventor-pair count $M_p$ (reflecting the costs/benefits of coordination and task specialization), and fixed effects, $f_p$, $\varphi_p$, and $\tau_p$, for firms, classes of International Patent Classification (IPC), and years, respectively.
$\varepsilon_p$ is a stochastic error.

\begin{comment}
Fig.\,\ref{fig:main}A shows the conditional expectation of $n_p$ obtained by estimating the model:
\begin{align}
n_{p}= \beta_0 &+ \beta_1 s_p + \beta_2 s_p^2 +  \cdots + \beta_m s_p^m \nonumber\\
    &+ \gamma \bar{K}_p+ \delta M_p + f_p + \varphi_p + \tau_p + \varepsilon_p\, \label{eq:regression}
\end{align}
where up to $m\,(\geq 1)$-th order polynomial terms of $s_p$ are considered.
The average cumulative-knowledge size of inventors in the patent team, $\bar{K}_p \equiv \frac{1}{H_p}\sum_{i\in {\bf H}_p} K_{it_p}$,
 controls individual ability in the team.
The inventor-pair count in the patent team, $M_p$, controls the team's potential costs and benefits of coordination and task specialization.
Fixed effects for firms, the subclass of International Patent Classification (IPC), and years are captured by $f_p$, $\varphi_p$, and $\tau_p$, respectively, and $\varepsilon_p$ is a stochastic error. The standard errors are clustered by IPC class.
\end{comment}

\begin{figure}[h!]
    \centering
    \includegraphics[width=0.9\columnwidth]{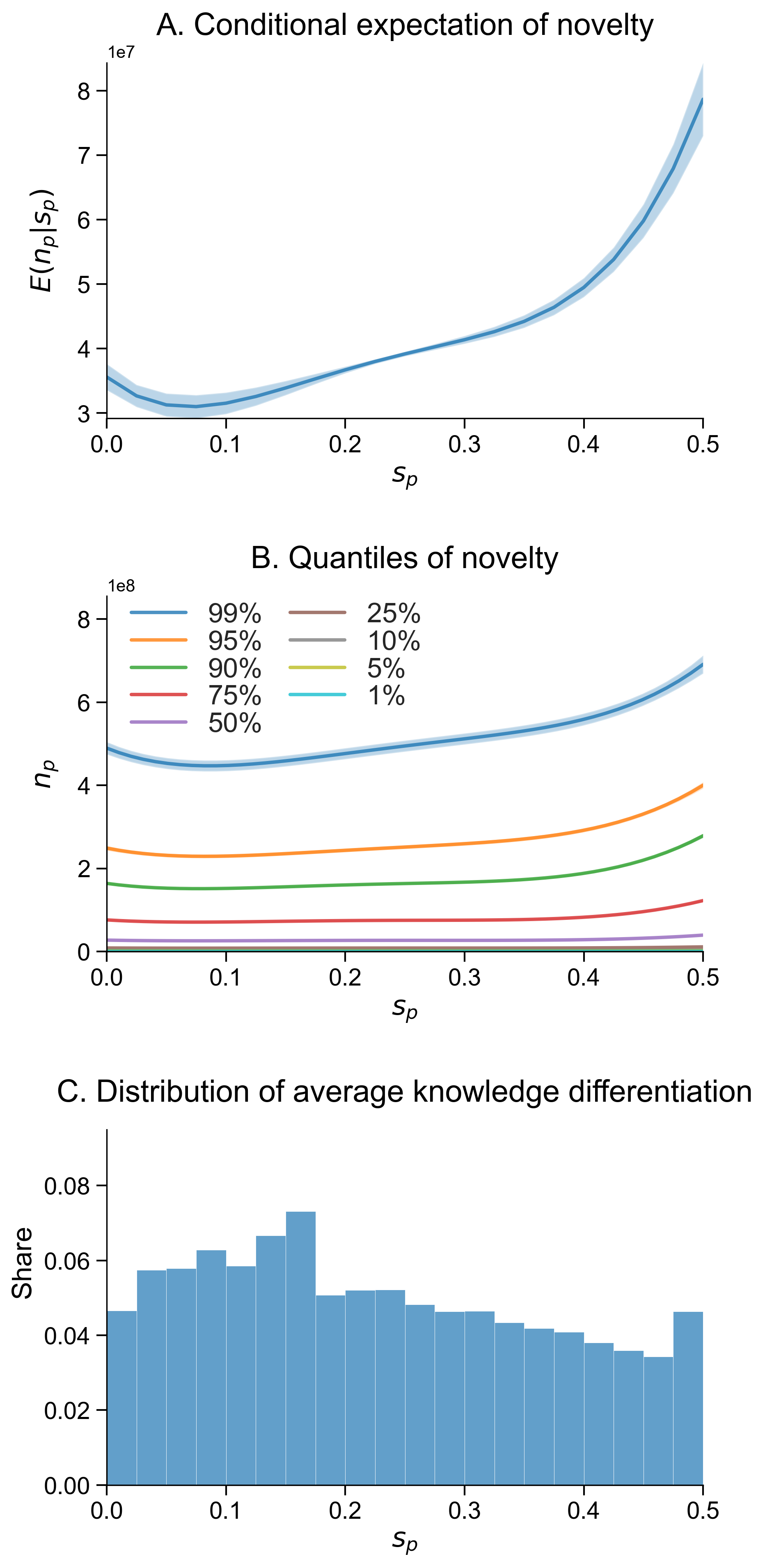}
    \caption{. (A,B) Estimated expectation and quantiles of $n_p$ conditional on $s_p$, respectively. 
    Other variables are evaluated at their mean. $m=4$ is chosen by the Bayesian Information Criterion for the regression to estimate $E(n_p | s_p)$.
    Shaded areas indicate 95\% confidence intervals. 
    In the regressions, standard errors are clustered by 124 IPC classes.
    The fixed effects $f_p$ are applied to firms that file 500 or more patents.     
    (C) Frequency distribution of $s_p\,(>0)$.}
    \label{fig:main}
\end{figure}

The estimated conditional expectation and quantiles of $n_p$ indicate a positive association between $s_p$ and $n_p$, except for a range of small $s_p$, while the observed $s_p$ are spread over the entire feasible range, $(0,0.5]$ (Fig.~\ref{fig:main}).

\begin{figure}[h!]
    \centering     \includegraphics[width=0.9\columnwidth]{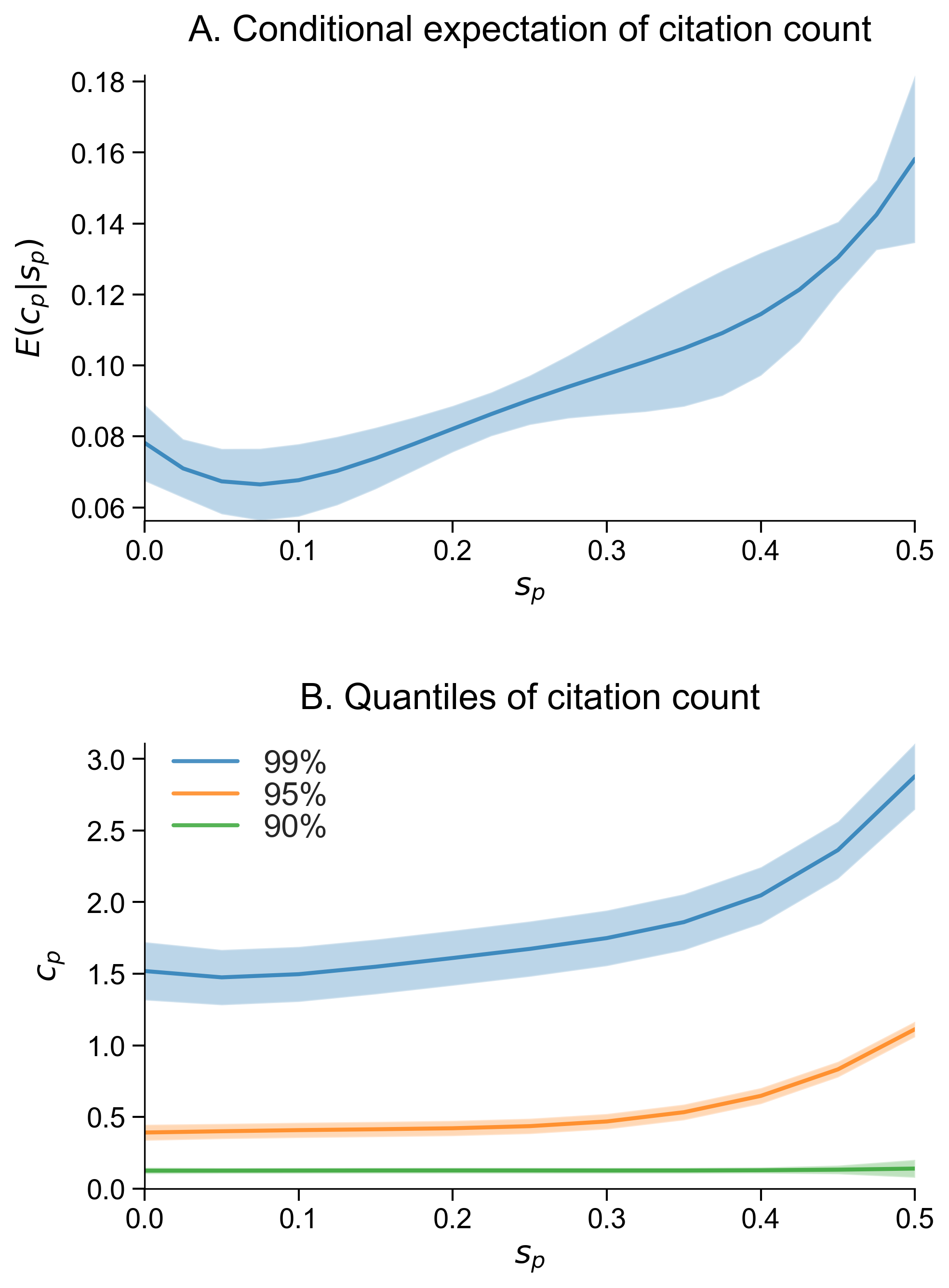}
    \caption{. (A,B) Estimated expectation and quantiles of $c_p$ conditional on $s_p$ corresponding to Fig.\,\ref{fig:main}(A,B), respectively. 
    $m=4$ is chosen by the Bayesian Information Criterion for the regression to estimate $E(c_p | s_p)$.
    Only the results for the 99, 95, and 90 percentiles are computed for Panel B since more than half the patents have zero citations.
    The shaded area indicates 95\% confidence interval.}
    \label{fig:citation}
\end{figure}

To see the robustness of the result, we  consider citation count of a patent as an alternative measure of output.
Let $\bar{c}_p$ be the citation count of patent $p$ within five years of application, excluding self-citations, where the self-citations include those by any patent $q$ whose set of inventors overlaps with that of $p$, that is, $\mathbf{H}_p \cap \mathbf{H}_q \neq \varnothing$.
We then use the citation count per inventor pair, $c_p = \bar{c}_p / M_p$, as an alternative output measure to $n_p$.
The basic relationships observed between $n_p$ and $s_p$ carry over to those between $c_p$ and $s_p$ as indicated by Fig.~\ref{fig:citation}A and B that correspond to Fig.~\ref{fig:main}A and B, respectively. 

%Column (2) of Table\,\ref{tab:regression} summarizes the estimation result of Eq.\,\ref{eq:regression} where the outcome variable is replaced by $c_p$.

\section*{Theoretical model and simulation}
There is a set of agents, $\mathbf{I}\equiv \{1,2,\ldots,I\}$, $I$ even, who collaborate in pairs to invent new knowledge. When agents $\{i,j\}$ collaborate on a given patent, the novelty $n$ of the patent is stochastic and follows a known probability distribution $P(n|s)$ conditional on the knowledge differentiation $s \in (0,0.5]$ between the collaborators. 
%This is essentially the distribution in Fig.~\ref{fig:main}B given $s_p$. 

The value $v=v(n)$ of an invention is an increasing function of the novelty of the invention. 
Assume that $v(\cdot)\ll \infty$ so that $V(s)\equiv E_{P(\cdot|s)}[v(n)]\ll \infty$.

The cost of collaboration by a pair $\{i,j\}$ is a continuous, increasing function $c_{ij}(s)$ of
%knowledge differentiation 
$s$ between the collaborators. Assume $c_{ij}(0) >0$. Collaboration carries a cost which increases as the knowledge sets of the collaborators become more differentiated, as there is less of a basis for the common understanding that assists effective communication. Various factors, such as geographic proximity and personality traits mean that this cost will be heterogeneous across pairs

As only profitable collaborations %will be
are pursued, the net value of a collaboration between %agents $i$ and $j$ will be
$\{i,j\}$ is given by $A_{ij} = \max\{0, V(s)-c_{ij}(s)\}$.
%on patent $p$  
For simplicity, assume that
$A_{ij}$ differs across all pairs $\{i,j\}$ and that
when a pair collaborates, $A_{ij}$ is split equally between members of a pair (see SI for an alternative value split by bargaining).
This defines a one-sided non-transferable utility matching problem \cite{GS1}.
Under these assumptions, a unique stable Pareto-efficient matching exists.
To find this matching, first match the pair $\{i,j\}$ with the largest $A_{ij}$ and remove them from $\mathbf{I}$.
Then, repeat this process until all agents match into collaborating pairs.

%\subsection*{Simulation}\:
To simulate a specific instantiation of this model, we consider technological categories with different levels of technological maturity that give rise to different levels of knowledge differentiation. Specifically, assume technological categories ${\bf  L}\equiv \{1,\ldots, L\}$ with each agent working in a single category. That is, $\mathbf{I}= \mathbf{I}^1 \cup \ldots \cup \mathbf{I}^L$, where $\mathbf{I}^l$ represents the set of agents working in category $l\in\mathbf{L}$; $\mathbf{I}^l \cap \mathbf{I}^m =\varnothing$ for $l\neq m$ . Assume that if agents $i$ and $j$ are in different categories, then $c_{ij}$ is large enough that such agents never collaborate.
%That is, $\mathbf{I}$ is partitioned into $\mathbf{I}^1,\ldots,\mathbf{I}^L$with $\mathbf{I}^l \cap \mathbf{I}^m =\varnothing\;\forall l\neq m$.

Let ${\bf K}^l \equiv \cup_{i\in {\bf I}^l} {\bf K}_i$ 
denote the knowledge set specific to category $l\in\mathbf{L}$.
Assume each agent $i$'s knowledge set has equal size $K_i = \bar{K} (< K^l\;\:\forall l\in \mathbf{L})$ and that all word pairs are equally novel so that $n_{wt}$ is the same for all $w$. Note that the maximum feasible value of $s_{ij}$ is increasing in $K^l$.
In this setup,  categories with smaller $K^l$ can be considered to represent more technologically \emph{mature} categories since agents have more knowledge in common, hence less room for knowledge recombination in these categories.

Assume that all categories share a value function, $v(n(s_{ij})) = \tilde{v}(s_{ij})e^{\varepsilon_{ij}}$, where $\tilde{v}(s)$ is given by the quartic function of $s$ from the estimated conditional expectation of the observed novelty of patents (Fig.~\ref{fig:main}A), with an appropriately adjusted intercept $v_0>0$. 
For agents within the same category, the cost function is given by $c_{ij} (s_{ij}) \equiv c(s_{ij})e^{\epsilon_{ij}}$ with $c(s)\equiv c_0 s$ and $c_0 > 0$.
$\varepsilon_{ij},\epsilon_{ij}\sim \mathcal{N}(0,1)$ are idiosyncratic noises.

\begin{figure}[h!]
    \centering
    \includegraphics[width=0.9\columnwidth]{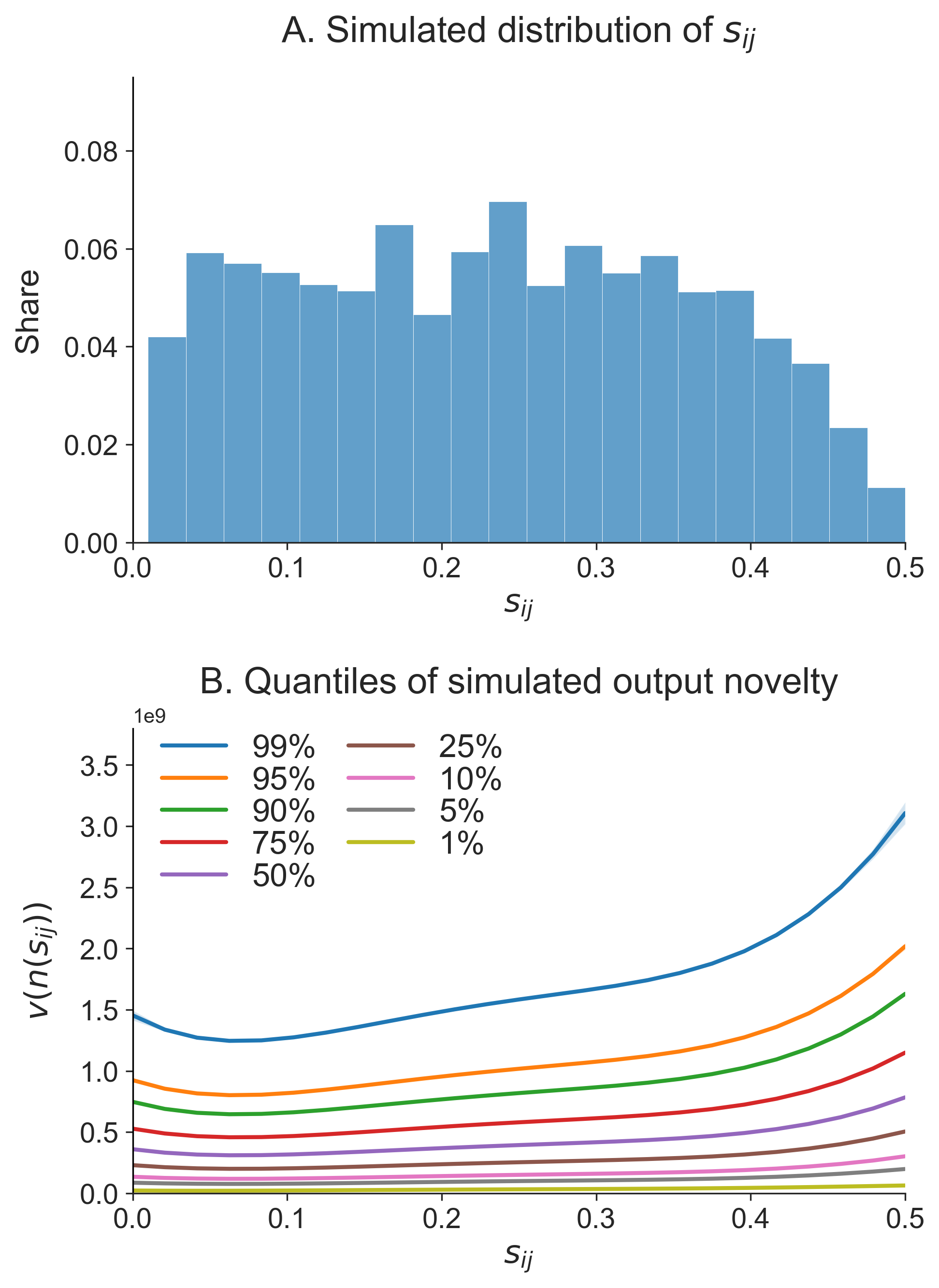}
    \caption{. Simulated pairwise collaboration.
    $L=100\text{,}000$, $I^l=100\;\:\forall l\in \mathbf{L}$, $\bar{K}=100$. $K^l$ $(l\in {\bf L})$ is given by an integer rounding the value drawn from Pareto distribution with scale and shape parameters being $\bar{K}+1$ and 2, respectively.
    $v_0=c_0=3.5\times 10^7$.
    (A) The distribution of $s_{ij}$ for matched inventor pairs.
    (B) Estimated quantiles of output values. Shaded areas indicate 95\% confidence intervals.}
    \label{fig:simulation}
\end{figure}

Fig.~\ref{fig:simulation} demonstrates that this model qualitatively replicates the observed variation in $s$ and in $n$ from inter-category variation in technological maturity and intra-category variation due to mismatch and idiosyncratic costs in collaboration.
The mismatch results from failing to achieve the best match due to the finiteness of the set of possible collaborators (see SI).

\section*{Discussion}
The current paper has found a direct, micro-level relationship linking knowledge exchange between collaborators to the novelty of patent applications. %Explicit consideration of knowledge differentiation is novel and crucial since
A team of inventors with diverse expertise does not necessarily promote knowledge exchange if the team members have collaborated in the past and the diverse knowledge is already their common knowledge \citep[cf.,][]{Singh-Fleming-MS2010,Singh-et-al-SMJ2016}.
Thus, team reshuffling is key to maintaining the novelty of collaborative outcomes \cite{Berliant-Fujita-IER2008}.  
%while maintaining sufficient common knowledge for smooth communication .

This provides support for a broader model class of knowledge spillovers in invention that have, thus far, lacked micro-level evidence \citep[e.g.,][]{Weitzman-QJE1998,Berliant-Fujita-IER2008}.
%This provides support for models of knowledge spillovers in invention that have, thus far, been expressed in the literature as a black box \cite{Weitzman-QJE1998, Zacchia-REStud2020}.
%
%Our findings of direct linkages between knowledge exchange between collaborators and the novelty of their outcome revealed a micro mechanism underlying the knowledge spillovers in invention that have, thus far, been expressed in the literature as a black box \cite{Weitzman-QJE1998, Zacchia-REStud2020}.
%
Importantly, our estimation shows a generally positive relationship between patent novelty and knowledge differentiation, with the highest novelty attained with the highest differentiation (Fig.\,\ref{fig:main}A). Our approach should also provide guidance for researchers studying different contexts involving knowledge exchange in collaboration \cite{bel2015team,newton2017shared} such as art creation \cite{uzzi2005collaboration} and community formation \cite{AngusNewton2015PLOS}.%, where this relationship might differ.

%Our findings are relevant to broader human activities that require collaborations, beyond research and development.
%DISCUSS ABOUT  COMPLEMENTARITY IN MORE DIVERSE LITERATURE.

\matmethods{%
\subsection*{Patent data} 
Patent data are taken from the \emph{published patent applications} of Japan \cite{Naito-Data}.
Identical inventors are traced by matching their names and the establishments they belong to.
%, recorded in the patent applications.
The data includes all patents filed between 1994 and 2017. We focus on those in 2009 and later,  970,197 of which involve multiple inventors.
The 1994-2008 data are used to compute the novelty of the patents filed after 2009.

\subsection*{Word standardization}
We use NLTK python library to standarize words as follows. (i) Nouns and verbs are lemmatized. %For example, ``plays'' and ``played'' are converted to ``play.''
(ii) Distinction between a verb and a noun is judged in the context. (iii) Numbers, non-alphabetical characters, and single-character words are removed. As an exception, hyphen-connected words (e.g., \emph{self-organization}) are kept. 
Noun/verb parts of these words are lammatized.
(iv) Stop-words (e.g., \emph{are} and \emph{also}) are removed.

\subsection*{Data availability} 
All data and codes are provided in SI. 

%TC:ignore
%\subsection*{Supporting Information}

%TC:endignore
}

\showmatmethods{} % Display the Materials and Methods section
%TC:ignore
\acknow{This research was conducted as part of the project, ``Agglomeration-based framework for empirical and policy analyses of regional economies,'' undertaken at the RIETI, Japan. 
This research has also been supported by the Kajima Foundation, and the Murata Science Foundation, the International Joint Research Center of Advanced Economic Theory of the Institute of Economic Research in Japan, and  the Grant in Aid for Research Nos.~17H00987, 19K15108, and 21H00695 of the MEXT, Japan. 
}
%\newpage

\label{page:lastpage}

\showacknow{}

% Display the acknowledgments section
% Bibliography
%TC:endignore
\bibliography{Collaboration}

%TC:ignore
%%% SI format %%%
\newpage
\setcounter{page}{1}
\setcounter{figure}{0}
\renewcommand{\thesection}{S\arabic{section}}
\renewcommand{\thesubsection}{S\arabic{section}.\arabic{subsection}}
\renewcommand{\thepage}{S\arabic{page}}
\renewcommand{\thefigure}{S\arabic{figure}}
\renewcommand{\theequation}{S\arabic{equation}}

\noindent {\bf\sffamily\Huge Supporting Information}

\section{Dataset}
Replication package including the datasets, Python and R codes is available from \url{https://www.dropbox.com/sh/ad1qrbkd5nik9x2/AAAOHuoQhO0hSkHoiUruKf3ha?dl=0}.

\section{Theoretical model and simulation}
\subsection*{An alternative matching with bargaining}
In the Model section it was assumed that value generated by collaborating agents is evenly split. This simplifies exposition, but no real problem arises if we assume that members of a pair can bargain over the allocation of surplus from a collaboration. In this case the matching problem become a one sided assignment game. There will still be Pareto efficient outcomes (matchings with allocation of surplus). However, these need not be stable. Indeed, the set of stable outcomes may be empty.

%\section{Simulation}
\subsection*{Optimal knowledge differentiation in the simulation model}
Assume that $K_i = K_j = \bar{K}$ for all agents, $i,j$, and that $n_{wt}=1$ for all $w$. This implies $K^D_{ij} \equiv k^D_{ij} = k^D_{ji}\equiv K^D_{ji}$ so that $s_{ij} = K^D_{ij}/K_{ij}$, where $K^D_{ij} = |{\bf K}^D_i \backslash {\bf K}^D_j|$ and $K_{ij} \equiv |
{\bf K}_i\cup {\bf K}_j|$.
Since the feasible values of $K^D_{ij}$ are $0,1,\ldots,\bar{K}$, the set of feasible values of $s_{ij}$ is given by
\begin{equation*}
    {\bf s}(\bar{K}) = \left\{\left.\frac{K^D}{K^D+\bar{K}}\;\right|\; K^D = 0, 1, \ldots, \bar{K}\right\}.
\end{equation*}

Optimal levels of knowledge differentiation between agents $i$ and $j$ for a given size, $\bar{K}$, of individual knowledge set are given by the set
\begin{equation*}
    s^*_{ij}(\bar{K}) = \arg \max_{s\in {\bf s}(\bar{K})} V(s)-c_{ij}(s).
\end{equation*}

Note that, even if $c_{ij}$ is identical across all pairs $\{i,j\}$, in a finite population it may still be impossible for a given agent $i$ to find and match with an agent $j$ satisfying $s_{ij} \in  s^*_{ij}(\bar{K})$. Such an agent may not exist in the population, or may have already been matched to someone else.
%TC:endignore

\end{document}